\documentclass[preprint,showpacs,preprintnumbers,amsmath,amssymb,superscriptaddress, prb]{revtex4}

\usepackage{graphicx}% Include figure files
\usepackage{dcolumn}% Align table columns on decimal point
\usepackage{bm}% bold math

\newcommand{\LAOLVO}[2]{LAO(#1)/LVO(#2)}

\begin{document}

\preprint{}

\title{Modulation Doping of a Mott Quantum Well by a Proximate Polar Discontinuity}% Force line breaks with \\

\author{T. Higuchi}
\email[E-mail: ]{higuchi@hwang.k.u-tokyo.ac.jp}
\affiliation{Department of Advanced Materials Science, 
University of Tokyo, Kashiwa, Chiba 277-8561, Japan}
\author{Y. Hotta}
\affiliation{Department of Advanced Materials Science, 
University of Tokyo, Kashiwa, Chiba 277-8561, Japan}
\author{T. Susaki}
\affiliation{Department of Advanced Materials Science, 
University of Tokyo, Kashiwa, Chiba 277-8561, Japan}
\author{A. Fujimori}
\affiliation{Department of Physics, 
University of Tokyo, Bunkyo-ku, Tokyo 113-0033, Japan}
\author{H. Y. Hwang}
\affiliation{Department of Advanced Materials Science, 
University of Tokyo, Kashiwa, Chiba 277-8561, Japan}
\affiliation{Japan Science and Technology Agency, Kawaguchi, 332-0012, Japan}


\date{\today}% It is always \today, today,

\begin{abstract}

We present evidence for hole injection into LaAlO$_3$/LaVO$_3$/LaAlO$_3$ quantum wells near 
a polar surface of LaAlO$_3$ (001). 
As the surface is brought in proximity to the LaVO$_3$ layer, 
an exponential drop in resistance 
and a decreasing positive Seebeck coefficient is observed 
below a characteristic coupling length of 10-15 unit cells. 
We attribute this behavior to a crossover 
from an atomic reconstruction of the AlO$_2$-terminated LaAlO$_3$ surface 
to an electronic reconstruction of the vanadium valence. 
These results suggest a general approach 
to tunable hole-doping in oxide thin film heterostructures.

\end{abstract}

\pacs{73.40.-c, 71.28.+d, 73.50.Lw, 71.30.+h}

\maketitle

\section{Introduction}

Surfaces and interfaces of oxides have been of
growing interest,
partially because of the rich variety of bulk oxide functionalities \cite{Imada},
as well as their unique reconstruction mechanisms \cite{Ohtomo04, Okamoto, Lee} 
not found in conventional semiconductors \cite{Baraff, Harrison}.
The observation of metallic interfaces between two perovskite insulators,
LaAlO$_3$ and SrTiO$_3$ \cite{Ohtomo04},
has motivated many studies
on the origin of this conductivity.
Two scenarios have been proposed, 
one based on electronic reconstructions driven by the polar discontinuity at the interface 
\cite{Nakagawa, Huijben, Thiel}, 
and another based on growth induced oxygen vacancies \cite{Kalabukhov, Siemons}.

The polar discontinuity scenario examines the built-in charge structure at the interface.
From an ionic point of view, LaAlO$_3$ is polar along the (001) direction 
with alternate stacking of
(LaO)$^+$ and (AlO$_2$)$^-$ layers, while SrTiO$_3$ is non-polar with
(SrO)$^0$ and (TiO$_2$)$^0$ layers.
When one unit cell (uc) of LaAlO$_3$ is placed on SrTiO$_3$,
there is a dipole shift in the electrostatic potential.
Additional LaAlO$_3$ layers build up this dipole shift,
leading to a diverging potential in the limit of infinite LaAlO$_3$ thickness.
To prevent this catastrophic situation, 
injection of  $-q/2$ (where $q$ is the elementary charge)
per 2D uc is needed at the 
(AlO$_2$)$^-$-(LaO)$^+$-(TiO$_2$)$^0$-(SrO)$^0$ interface,
which can be accommodated by a partial valence change
of Ti$^{4+}$ to Ti$^{3+}$ near the interface.
The Ti$^{3+}$ component provides mobile SrTiO$_3$ conduction electrons
in this picture \cite{Nakagawa}.
Alternatively, it has been proposed 
that the essential origin of the conducting interface is 
the formation of SrTiO$_3$ surface oxygen vacancies during the highly kinetic growth
of films by pulsed laser deposition (PLD) \cite{Siemons, Kalabukhov}.
SrTiO$_3$ is known to be a material
which easily accommodates oxygen vacancies
that readily dope itinerant electrons \cite{Muller06}.
Discriminating between these two proposed scenarios has been controversial,
in part because both mechanisms could give similar transport and spectroscopic signatures.

The LaAlO$_3$/SrTiO$_3$ system actually involves two polar discontinuities 
- the interface just described, as well as the polar AlO$_2$-terminated surface of LaAlO$_3$, 
which requires a net $+q/2$ per 2D uc.
Recently it was found that the conducting LaAlO$_3$/SrTiO$_3$ interface
exhibited a transition to an insulating state
when the LaAlO$_3$ was thinner than a critical thickness of 4 uc, 
bringing the two polar discontinuities close together \cite{Thiel}.
(Similar behavior was also observed for proximity coupling
of two LaAlO$_3$/SrTiO$_3$ interface polar discontinuities with opposite sign \cite{Huijben}.)
This critical thickness may be interpreted as the threshold dipole shift
below which it is energetically favorable to remain
in an unreconstructed state for both the surface and the interface \cite{HYH06}.
A different but equivalent perspective is that
the polar AlO$_2$-terminated surface of LaAlO$_3$ is compensating
the electrons at the LaAlO$_3$/SrTiO$_3$ interface by hole-doping on short length scales.
Far above the critical thickness, the polar LaAlO$_3$ (001) surface
atomically reconstructs via surface off-stoichiometry and surface relaxation;
electronic reconstructions are unavailable
because of the fixed valence of La, Al, and O \cite{Yao, Francis, Lanier}.
Below a characteristic coupling distance, however,
hole-doping provides an alternative electronic reconstruction
if it is energetically favorable.

In order to test this possibility of hole-doping, 
we have studied the transport properties of the Mott insulator LaVO$_3$ 
embedded in LaAlO$_3$ in trilayer structures [Fig.~\ref{cap-fig1}(a)].
LaVO$_3$ is an attractive candidate 
since it can be readily hole-doped by chemical substitution  \cite{Miyasaka},
and because of its structural and thermodynamic compatibility 
with LaAlO$_3$ for the growth of atomically precise thin film structures.
Although LaVO$_3$ has the same polar structure as LaAlO$_3$,
and hence no polar discontinuity at their interface, 
we find an exponential drop in resistance 
when a polar AlO$_2$-terminated surface of LaAlO$_3$ is brought 
in close proximity to the LaVO$_3$ quantum well.
Furthermore, the positive thermopower voltage measured indicates hole-like carriers,
which scales with the doping dependence of bulk LaVO$_3$.
These results indicate that polar discontinuities can be 
utilized for the tunable doping of holes, 
which cannot arise by growth induced oxygen vacancies.

\section{Methods}

\begin{figure}[]
\includegraphics{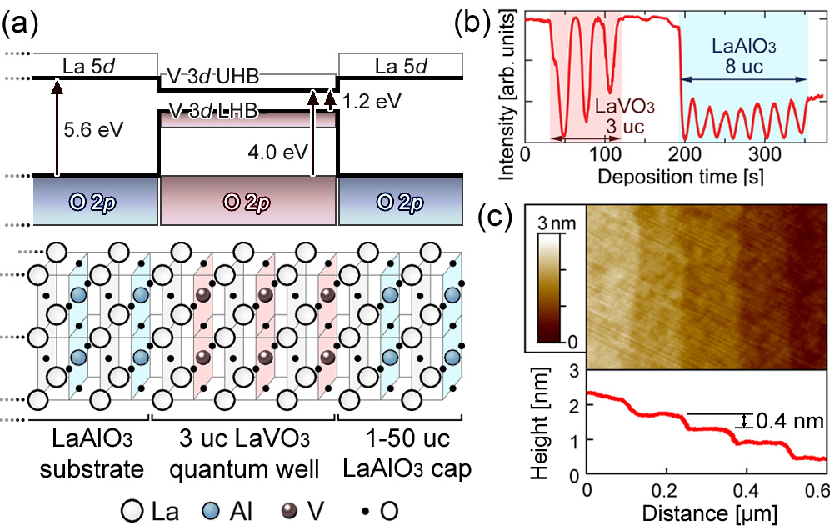}% Here is how to import EPS art
\caption{\label{cap-fig1} (Color online)
(a) Schematic band diagram and crystal structure of \LAOLVO{$n$}{3}
quantum wells grown on AlO$_2$-terminated LaAlO$_3$ (001) substrates. 
Filled and empty bands correspond to the valence and conduction bands, respectively,
and O 2$p$ bands are assumed to be aligned.
(b) Typical RHEED oscillations during growth of \LAOLVO{8}{3}.
(c) AFM image of \LAOLVO{50}{3} showing a clear step and terrace surface
with step height of $\sim $ 0.4 nm.}
\end{figure}

LaAlO$_3$ ($n$ uc)/LaVO$_3$ ($m$ uc)/LaAlO$_3$ (substrate) heterostructures 
(\LAOLVO{$n$}{$m$}) 
were fabricated by PLD, using a KrF excimer laser with a laser fluence of 1.0 J/cm$^2$, 
a spot size of 1.6 mm$^2$, and repetition rate of 8 Hz.
The thickness of each layer was monitored 
by reflection high-energy electron diffraction (RHEED) oscillations [Fig.~\ref{cap-fig1}(b)].
AlO$_2$-terminated LaAlO$_3$ (001) single crystal substrates were preannealed at 950 $^{\circ}$C for 30 minutes, 
and $m$ uc ($m$ = 1 to 30) LaVO$_3$ layers were deposited, 
followed by $n$ uc ($n$ = 1 to 50) LaAlO$_3$ capping layers 
using LaVO$_4$ polycrystalline and LaAlO$_3$ single crystal targets, respectively. 
The growth temperature was 600 $^{\circ}$C and the oxygen partial pressure was 1.0 $\times$ $10^{-6}$ Torr for all processes. 
These conditions follow our previous optimization of high quality LaVO$_3$ thin film growth 
in the layer-by-layer growth mode \cite{Hotta06g}, with one exception. 
Note the different laser conditions: 
we now use a 4 lens afocal zoom stage to accurately image an aperture, 
rather than a single lens just off of the focusing condition. 
This is far less sensitive to the divergence characteristics of the laser, 
and hence much more reproducible between systems. 
We confirmed that one RHEED oscillation corresponds to the formation of one perovskite unit cell 
by characterizing superlattices of LaAlO$_3$ and LaVO$_3$ using x-ray diffraction.

The stoichiometry of LaVO$_3$ films grown in these conditions has been confirmed 
from several different perspectives in previous studies. 
The film lattice volume was found to be close to bulk (accounting for the compressive strain by the substrate) 
-- usually, in the presence of significant cation defects, there is significant lattice expansion of the lattice \cite{Hotta06g}. 
LaVO$_3$ films were also studied extensively 
by scanning transmission electron microscopy and electron energy-loss spectroscopy (EELS) \cite{Fitting06}. 
The insulating LaVO$_3$ films have a clear V$^{3+}$ valence, 
as observed in the V-L$_{2,3}$ edge. 
Any significant doping by off-stoichiometry would have shown clear deviations from V$^{3+}$ spectra. 
Thus the stoichiometry of our LaVO$_3$ films has been well established in these previous studies.
LaAlO$_3$/LaVO$_3$/LaAlO$_3$ quantum wells could not be grown at higher oxygen partial pressures than used here, 
due to the competing formation of polycrystalline LaV$^{5+}$O$_4$ \cite{Hotta06g, Fitting06}. 
Furthermore, the LaVO$_3$ layers in the as-grown quantum wells converted to the insulating, 
transparent ($d^0$) LaV$^{5+}$O$_4$ phase upon oxygen post annealing.

The film surface was investigated by atomic force microscopy (AFM),
and all structures were atomically flat 
with a clear step and terrace structure
reflecting the slight miscut angle of the substrate [Fig.~\ref{cap-fig1}(c)].
The step size ($\sim$ 0.4 nm) was consistent with the height of one LaAlO$_3$ uc
(pseudocubic lattice constant $a_{\mathrm{LaAlO_3}}=0.379$ nm).
Ohmic contacts were made to the buried quantum well layers
using indium ultrasonic soldering, which penetrated more than 40 nm from the surface
as confirmed using buried SrVO$_3$ test structures.

\section{Results and Discussion}

\subsection{LaAlO$_3$ cap thickness dependence}

\begin{figure}[]
\includegraphics{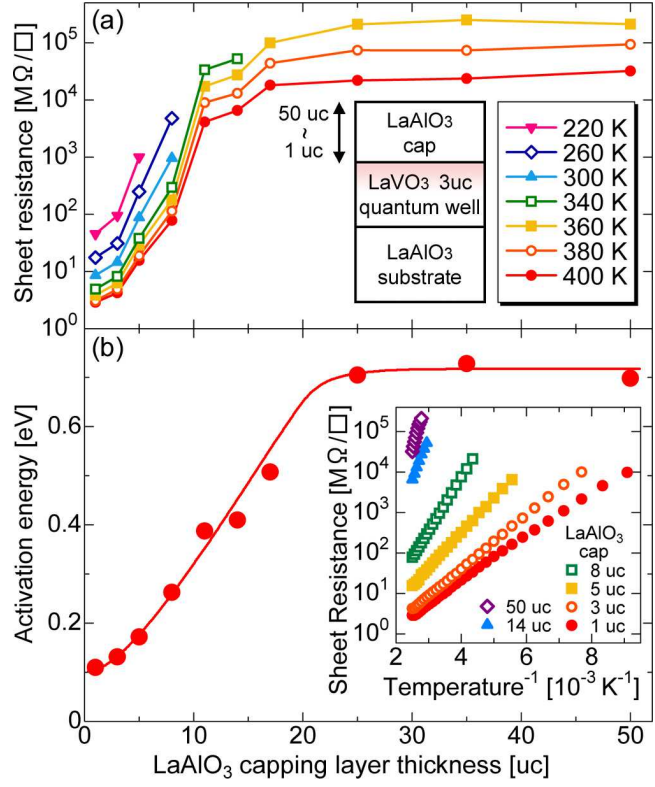}% Here is how to import EPS art
\caption{\label{cap-fig2} (Color online)
(a) Sheet resistance and (b) activation energy of \LAOLVO{$n$}{3}
as a function of LaAlO$_3$ cap thickness $n$. The curve is a guide to the eye.
Inset of (b) shows Arrhenius plots of the sheet resistance.
}
\end{figure}

The in-plane sheet resistance
was dramatically dependent on the LaAlO$_3$ cap thickness
as shown in Fig.~\ref{cap-fig2}(a), 
where the thickness of the LaVO$_3$ layer was fixed to 3 uc.
Below a characteristic thickness of around 10-15 uc,
the sheet resistance decreased exponentially.
For all samples the sheet resistance showed Arrhenius-type thermally activated behavior
as shown in the inset of Fig.~\ref{cap-fig2}(b).
The activation energy was 
$\sim 0.7$ eV for the thick capping layer samples,
and decreased to $\sim 0.1$ eV
with a similar characteristic thickness.
Since LaAlO$_3$ is such a robust insulator 
(both in bulk and as measured in our thin films), 
the resistance of the embedded LaVO$_3$ layer could be measured 
to the very high values shown.

Similar results were found for these \LAOLVO{$n$}{$3$} trilayer structures 
further capped with a non-polar material, 10 uc of SrTiO$_3$,
but measurement was limited to a lower range of resistance values ($n \leq 8$) 
by the higher intrinsic conductivity of SrTiO$_3$.
This suggests that the conductivity of the LaVO$_3$ layer 
only depends on the distance to the polar surface (LaAlO$_3$ surface) or interface (SrTiO$_3$/LaAlO$_3$),
not on the total thickness of the material deposited on top of it.
Therefore, the role of LaVO$_3$ defects created during growth of the cap can be neglected here.

\begin{figure}[]
\includegraphics{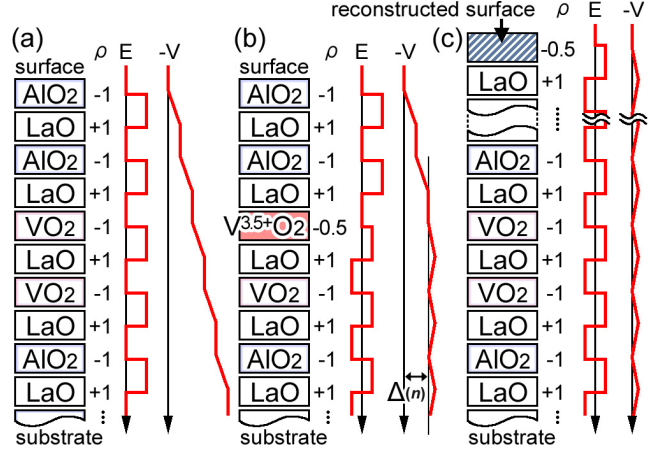}% Here is how to import EPS art
\caption{\label{cap-fig3} (Color online)
Schematic diagrams of possible reconstruction processes.
(a) Without reconstruction 
the structure is composed
of negatively and positively charged ($\rho$) layers from the surface,
which induces a non-negative electric field ($E$),
leading to a divergence in potential ($-V$).
(b) Electronic reconstruction at the LaVO$_3$ quantum well layer
with a net half hole per 2D unit cell induced.
The potential divergence is canceled, but a finite shift $\Delta(n)$ in potential remains.
(c) Atomic reconstruction of the LaAlO$_3$ polar surface.
}
\end{figure}

These results are well explained by the electrostatic coupling of 
reconstructions of the polar surface
and the quantum well, as illustrated in Fig.~\ref{cap-fig3}.
Since we fabricated the quantum well structures 
by perovskite unit cell deposition
on AlO$_2$-terminated LaAlO$_3$ substrates,
the surface of the LaAlO$_3$ cap preserves AlO$_2$-termination,
which requires a net $+q/2$ per 2D uc to avoid the potential divergence
arising from the surface polar discontinuity.
Here we consider two reconstruction mechanisms for this system.
The first possibility is hole injection into the LaVO$_3$ quantum well layer.
In this case we have no divergence, 
but still a finite dipole shift $\Delta(n)$ arising from the polar LaAlO$_3$ cap.
Thus the total energy of this electronic reconstruction is the sum of
the energy cost to change the vanadium valence $E_{val}$ and $\Delta(n)$.
The second process is the normal surface reconstruction of LaAlO$_3$
where an atomic reconstruction  
is dominant (oxygen vacancies and lattice distortions) 
to provide positive charge.
This reconstruction requires an energy cost $E_{sur}$
which is nominally independent of the thickness of the cap layer. 
Note that this is the same $E_{sur}$, calculated in Ref. [\onlinecite{Levy}] with respect to $E_{val}$ for Ti, 
where the conducting LaAlO$_3$/SrTiO$_3$ interface is toggled by manipulating the surface reconstruction.
The difference in $E_{val}$ for Ti and V leads to the different critical thickness observed in the two systems.

In the simplest form, $\Delta(n)$ increases linearly as a function of the LaAlO$_3$ cap thickness:
$\Delta(n)=n q / 2 \varepsilon a_{\mathrm{LaAlO_3}} $ 
($\varepsilon$ is the dielectric constant of LaAlO$_3$).
If the cap is sufficiently thin (and $E_{val} < E_{sur}$), 
then $E_{val} + \Delta(n) < E_{sur}$ 
and hole injection is dominant, which decreases the resistance of the quantum well.
On the other hand, when the cap is very thick, $E_{val} + \Delta(n) > E_{sur}$,
and the surface of the LaAlO$_3$ cap is reconstructed, 
while the LaVO$_3$ layers are undoped and insulating.
Between these limiting cases, 
hole injection into the LaVO$_3$ layer gradually decreases
and the resistance increases
as a function of the cap thickness,
as observed in the transport measurements.
Taking the LaAlO$_3$ bulk dielectric constant $\varepsilon = 24 \varepsilon_0$
($\varepsilon_0$ is the vacuum permittivity),
$\Delta(10)= 9.9$ eV, exceeding the LaAlO$_3$ bulk bandgap $\sim 5.6$ eV.
In the actual system this is greatly reduced by strong polarization of the LaAlO$_3$ lattice
as observed by surface x-ray diffraction in LaAlO$_3$ thin films on SrTiO$_3$ \cite{Vonk}.
Thus the detailed response is far more complex than represented in Fig.~\ref{cap-fig3},
and the energy cost $\Delta (n)$ is likely dominated by the polarization energy of LaAlO$_3$.
A further discussion of the contrast in LaAlO$_3$ thickness dependence of the conductivity observed here 
(conductivity in the thin limit), with prior results for the (001)-oriented LaAlO$_3$/SrTiO$_3$ 
and LaVO$_3$/SrTiO$_3$ interfaces \cite{Thiel, Hotta07} (conductivity in the thick limit), is given in the Appendix.

\begin{figure}[]
\includegraphics[width=3.18cm]{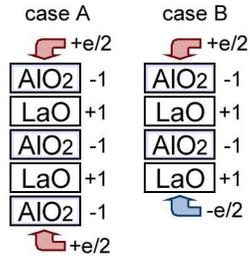}% Here is how to import EPS art
\caption{\label{cap-fig4} (Color online) Schematic reconstructions of the top and bottom polar surfaces 
of (001)-oriented LaAlO$_3$ for two different terminations of the bottom surface.
}
\end{figure}

Note that Fig.~\ref{cap-fig3} only shows the near surface region with the LaVO$_3$ quantum well. 
To formally confirm that global charge neutrality is maintained for the total system in all cases of reconstructions, 
the structure of the bottom surface must be known. 
This can best be illustrated in the schematic shown in Fig.~\ref{cap-fig4}, 
where we discuss only polar surfaces of LaAlO$_3$ for simplicity.

As is well established in surface science, 
the polar surfaces are unstable to reconstructions driven to keep the electrostatic potential bounded. 
For case A, each surface requires $+e/2$ charge to achieve this. 
This net $+e$ is compensated by the extra (AlO$_2$)$^-$ layer in the total structure, thus preserving global charge neutrality. 
For case B, the top surface requires $+e/2$ charge, 
the bottom surface $-e/2$ charge, and since the number of (AlO$_2$)$^-$ layers and (LaO)$^+$ layers are equal, 
here again global charge neutrality is conserved. 
Note that in case B, one could say that the $e/2$ was transferred from the bottom to the top surface, 
while in case A, $e/2$ may be considered to arise locally. 
In either case, the reconstruction of the top (AlO$_2$)$^-$ surface is identical, 
although to formally conserve total charge, knowledge of the bottom interface is needed. 
Practically, these two surfaces are completely decoupled (the thickness of our substrates are 0.5 mm, 
and certainly changing the termination layer on one surface does not affect the other across this macroscopic distance), 
so we address only the top surface in Fig.~\ref{cap-fig3}, as is conventional in such discussions of polar surfaces. 
Therefore our conclusions are independent of the assignment of charge transfer. 
Only when polar surfaces are microscopically close \cite{Huijben, Thiel, Hotta07} do they couple.

\subsection{Thermoelectric power}

\begin{figure}[]
\includegraphics[width=8.6cm]{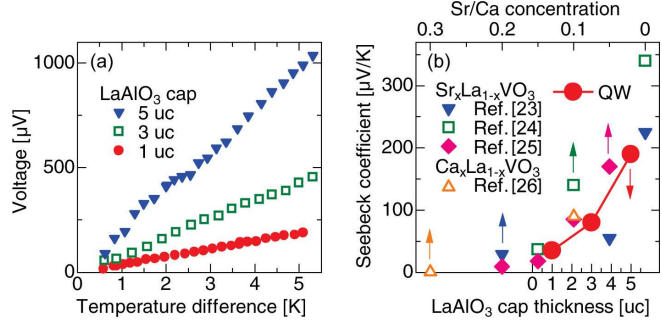}% Here is how to import EPS art
\caption{\label{cap-fig5} (Color online)
(a) Thermoelectric voltage measured as a function of applied temperature difference
between two electrodes on \LAOLVO{$n$}{3} quantum wells
at room temperature.
(b) Seebeck coefficient of \LAOLVO{$n$}{3} as a function of $n$ (filled circles), plotted in comparison with bulk values for La$_{1-x}$Sr$_x$VO$_3$ (Refs. [\onlinecite{Sayer, Webb, Dougier}]) and
La$_{1-x}$Ca$_x$VO$_3$ (Ref. [\onlinecite{Nguyen}]) at 300 K.
}
\end{figure}

A critical test of this model of coupled surface and interface reconstructions 
would be whether the carriers induced by the electronic reconstruction are holes, 
not electrons as in all previous examples 
\cite{Ohtomo04, Nakagawa, Huijben, Thiel, Kalabukhov, Siemons}.
They should further be tunable with the LaAlO$_3$ cap layer thickness.
To determine the sign of the transport carriers, 
the Seebeck coefficient $S$ of the \LAOLVO{$n$}{3} structures was measured, 
as shown in Fig.~\ref{cap-fig5}.
Intrinsic voltage fluctuations at high impedance limited the measurements
to higher temperatures and thin LaAlO$_3$ cap samples.
The positive sign of $S$ confirms hole-doping, and it systematically 
increased as a function of increasing LaAlO$_3$ capping layer thickness,
indicating a decreasing hole density and 
mirroring the evolution of the sheet resistance shown in Fig.~\ref{cap-fig2}.
These results agree well with thermopower measurements
of bulk La$_{1-x}$Sr$_x$VO$_3$ and La$_{1-x}$Ca$_x$VO$_3$,
where $S$ was positive for $0<x<0.2$ and
decreased as the dopant concentration $x$ increased \cite{Sayer, Webb, Dougier, Nguyen}.  
Given the compressive strain arising from the LaAlO$_3$ substrate, 
the reduced electronic bandwidth, 
and interface scattering, 
a comparison between these LaVO$_3$ quantum wells and bulk values is an approximate one. 
Nevertheless, this comparison indicates 
that the maximum equivalent hole density achieved 
in the single uc LaAlO$_3$ cap sample is just below the bulk metal-insulator transition, 
occurring at $x=0.18$ in La$_{1-x}$Sr$_x$VO$_3$ \cite{Miyasaka}.
A recent theoretical proposal suggests 
that doped holes at LaVO$_3$ interfaces are susceptible to charge ordering due to the artificial confined geometry 
\cite{Jackeli}, which may be relevant here.

\subsection {LaVO$_3$ thickness dependence}

\begin{figure}[]
\includegraphics[width=8.6cm]{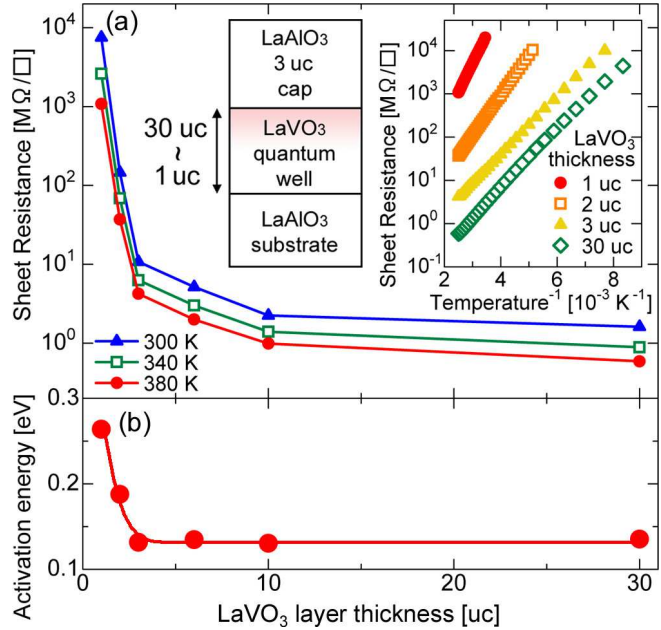}% Here is how to import EPS art
\caption{\label{cap-fig6} (Color online)
(a) Sheet resistance and (b) activation energy of \LAOLVO{3}{$m$} 
as functions of LaVO$_3$ layer thickness $m$. Inset of (a) shows Arrhenius plots of the sheet resistance. 
The curve is a guide to the eye.}
\end{figure}

In the regime of a thin LaAlO$_3$ cap layer, 
a V$^{4+}$ component giving rise to these holes should be observable.  
This has indeed been seen recently by x-ray photoemission spectroscopy 
on \LAOLVO{3}{$m$} structures (grown on SrTiO$_3$ substrates) \cite{Hotta06}.  
In particular, the spatial distribution was highly asymmetric, 
with the V$^{4+}$ predominantly in the topmost LaVO$_3$ layer.  
This non-uniformity naturally arises in Fig.~\ref{cap-fig3}(b), 
since $\Delta(n)$ would thus be minimized.  
The detailed charge distribution is a balance between $\Delta(n)$ and 
the electronic compressibility.
This issue was further studied
by measuring the transport properties of \LAOLVO{3}{$m$}
quantum well structures.
As shown in Fig.~\ref{cap-fig6},
the sheet resistance was strongly dependent on the LaVO$_3$ layer thickness $m$.
The data did not scale with $1/m$,
as would be expected for a uniform 3D resistivity.
For $m>10$, there was little change in the sheet resistance, but 
for $m<6$ it increased more rapidly than $1/m$,
indicating the length scale for strong perturbation of the charge distribution 
by confinement effects. 
Thus the thickness dependent resistance indicates 
that the conducting holes were distributed primarily at the top of the LaVO$_3$ layer,
as would be expected for doping arising from the polar LaAlO$_3$ surface.

\section{Summary}

We have found a strong electrostatic coupling 
between the AlO$_2$-terminated LaAlO$_3$ (001) surface 
and an embedded LaVO$_3$ quantum well. 
When they are separated by less than 4-6 nm, 
transport measurements indicate systematically increasing hole-doping 
with decreasing separation. 
We propose that these results reflect a competition 
between atomic and electronic reconstructions, 
driven by the need to resolve the divergent surface energy 
arising from the polar surface termination. 
An important aspect of this study is 
that the electronic reconstruction involves holes, 
not electrons. 
Therefore the role of polar discontinuities 
can be more clearly distinguished from that of electron-donor oxygen vacancies, 
in contrast to previous examples. 
They further demonstrate that oxide heterostructures 
can be designed to introduce carriers 
without local chemical substitution, 
in analogy to modulation doping in compound semiconductor heterostructures.

\section*{Acknowledgments}

We thank Y. Hikita, K. Itaka, and T. Higuchi (Sr.) for
helpful discussions. We acknowledge support from a
Grant-in-Aid for Scientific Research on Priority Areas.
Y. H. acknowledges partial support from QPEC, Graduate
School of Engineering, University of Tokyo.

\section*{Appendix}

\subsection*{Comparison of LaAlO$_3$ Thickness-Dependent Conductivity in LaAlO$_3$/LaVO$_3$/LaAlO$_3$  (001)
with LaAlO$_3$/SrTiO$_3$ (001) and LaVO$_3$/SrTiO$_3$ (001)}

\begin{figure}[]
\includegraphics[width=8.6cm]{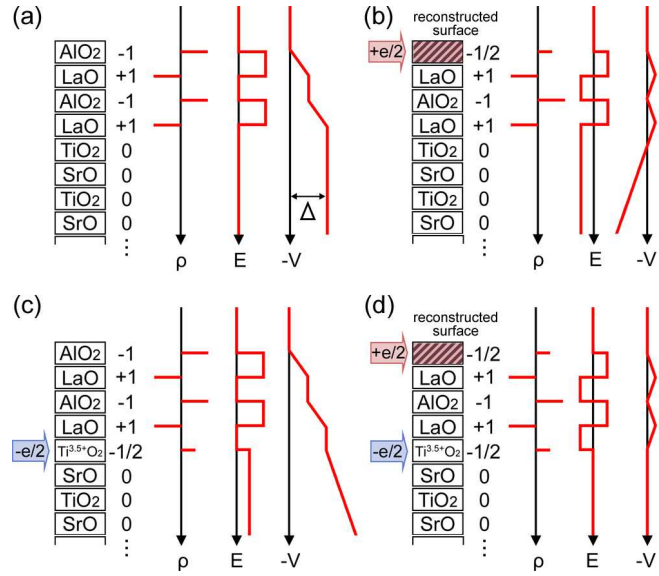}% Here is how to import EPS art
\caption{\label{cap-fig7} (Color online)
Schematic reconstructions of the polar surface of (001)-oriented LaAlO$_3$ and the LaAlO$_3$/SrTiO$_3$ interface. 
(a) No reconstruction, (b) only the LaAlO$_3$ surface is reconstructed, 
(c) only the LaAlO$_3$/SrTiO$_3$ interface is reconstructed, and (d) both are reconstructed.}
\end{figure}

It is worth contrasting the LaAlO$_3$ thickness dependence for the interface hole-doping we find here, 
with prior results for electron-doping at the (001)-oriented LaAlO$_3$/SrTiO$_3$ interface \cite{Thiel} 
(and (001)-oriented LaVO$_3$/SrTiO$_3$ interface \cite{Hotta07}). 
For the LaAlO$_3$/SrTiO$_3$ structures, there are two polar discontinuities, 
the LaAlO$_3$ surface and the LaAlO$_3$/SrTiO$_3$ interface. 
Without any reconstruction, the finite shift $\Delta$ remains, as shown in Fig.~\ref{cap-fig7}(a). 
When the LaAlO$_3$ layer is thick, the large $\Delta$ makes the system unstable, 
and the polar discontinuities are reconstructed in order to solve the instability -- 
the LaAlO$_3$ surface as shown in Fig.~\ref{cap-fig7}(b), 
and the LaAlO$_3$/SrTiO$_3$ interface by inducing Ti valence changes 
and associated metallic behavior as shown in Fig.~\ref{cap-fig7}(c).
However, one reconstruction just by the LaAlO$_3$ surface or by the LaAlO$_3$/SrTiO$_3$ interface alone is insufficient 
and the electrostatic potential diverges. 
Therefore, when the two polar discontinuities are far apart from each other 
and the potential shift $\Delta$ in Fig.~\ref{cap-fig7}(a) becomes large, 
both must reconstruct simultaneously, as shown in Fig.~\ref{cap-fig7}(d). 
Only when they are brought close together can they couple and remain unreconstructed \cite{Huijben, Thiel, Hotta07}. 
Thus the LaAlO$_3$/SrTiO$_3$ interface is insulating for thin LaAlO$_3$ layers [Fig.~\ref{cap-fig7}(a)], 
and metallic in the thick limit [Fig.~\ref{cap-fig7}(d)].

For the LaAlO$_3$/LaVO$_3$/LaAlO$_3$ structures, there is only one polar discontinuity, the polar LaAlO$_3$ surface. For a LaVO$_3$ layer embedded in LaAlO$_3$ far from any surface, there are no discontinuities to resolve, and thus no reconstructions induced. Only when the LaVO$_3$ layer is brought in proximity to the polar LaAlO$_3$ surface does it provide an alternative reconstruction mechanism, by V valence changes. Therefore conductivity appears only in the thin limit, opposite to that for the LaAlO$_3$/SrTiO$_3$ interface.

\newpage

\end{document}